\documentclass{article}
\usepackage{frascatiphys,here,graphicx,subfigure}
\begin{document}
\title{ 
Higgs bosons in $t\bar{t}$ production}
\author{
Roberto Barcel\'o    \\
{\em CAFPE and Departamento de F{\'\i}sica Te\'orica y del Cosmos,} \\
{\em Universidad de Granada, E-18071 Granada, Spain.} \\
} 
\maketitle 
\baselineskip=11.6pt
\begin{abstract}
The top quark has a large Yukawa coupling with the 
Higgs boson. In the usual extensions of the standard model the
Higgs sector includes extra scalars, which also tend to couple 
strongly with the top quark. Unlike the Higgs, these fields 
have a {\it natural} mass above $2m_t$, so they could introduce
anomalies in $t \bar t$ production at the LHC. We study their
effect on the $t \bar t$ invariant mass distribution 
at $\sqrt{s}=7$ TeV. We focus on the bosons ($H$,$A$) of
the minimal SUSY model and on the scalar field ($r$) associated to
the new scale $f$ in Little Higgs (LH) models. We show that in all 
cases the interference with the standard amplitude dominates over the 
narrow-width contribution. As a consequence, the
mass difference between $H$ and $A$ or the
contribution of an extra $T$-quark loop in LH models become
important effects in order to determine if 
these fields are observable there. We find that a 1 fb$^{-1}$
luminosity could probe the region $\tan\beta \le 3$ of SUSY
and $v/(\sqrt{2}f) \ge 0.3$ in LH models.
\end{abstract}
\baselineskip=14pt
\section{Introduction}
The main objective of the LHC is to reveal the nature of the
mechanism breaking the electroweak symmetry. This requires
not only a determination of the Higgs mass and couplings,
but also a search for additional particles that may be 
related to new dynamics or symmetries present
at the TeV scale. 
The top-quark sector appears then as a promising place 
to start the search, as it is there where the EW symmetry 
is broken the most. Generically, the large top-quark 
Yukawa coupling with the Higgs boson ($h$) also 
implies large couplings with the extra physics. For example, in
SUSY extensions $h$ comes together with neutral scalar ($H$) 
and pseudoscalar ($A$) fields\cite{Djouadi:2005gj}. 
Or in Little Higgs (LH) models, a global symmetry in the Higgs 
and the top-quark sectors introduces
a scalar singlet and an extra $T$ quark\cite{Schmaltz:2005ky,Perelstein:2005ka}. 
In all cases these scalar fields have large Yukawa couplings that could
imply a sizeable production rate in hadron collisions and a
dominant decay channel into $t\bar t$. 
\section{Top quarks from scalar Higgs bosons}
\begin{figure}[H]
    \begin{center}
        {\includegraphics[scale=0.4]{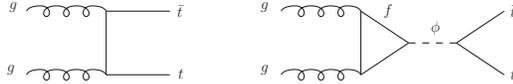}}
        \caption{\it Diagrams that interfere in $t\bar t$ production.}
\label{fig1}
    \end{center}
\end{figure}

The potential to observe new physics in $m_{t\bar t}$
at hadron colliders has been discussed in 
previous literature\cite{Dicus:1994bm,Frederix:2007gi}. 
In general, any heavy $s$--channel 
resonance with a significant
branching ratio to $t\bar t$ will introduce distortions. 
In the diagram depicted in fig.\ref{fig1} the intermediate scalar is produced 
at one loop, but the gauge and Yukawa couplings are all strong.
In\cite{Barcelo:2010bm} we give the expressions for the leading-order 
differential cross section for $gg\rightarrow t\bar t$ through a scalar 
and a pseudoscalar, $\phi$. To have an observable effect it is essential that the width
$\Gamma_\phi$ is small. This is precisely the
reason why the effect on $m_{t\bar t}$
of a very heavy standard Higgs $h$ would
be irrelevant. A 500 GeV Higgs boson would couple strongly
to the top quark, but even stronger to itself.
Its decay into would-be Goldstone bosons would then dominate, implying a 
total decay width of around 60 GeV.

To have a smaller width and a larger effect
the mass of the resonance must {\it not} be EW.
In particular, SUSY or LH models
provide a new scale and massive Higgses with no need for large
scalar self-couplings.
\section{SUSY neutral bosons}
SUSY incorporates two Higgs doublets, and after EWSB there are two neutral bosons ($H$ and
$A$) in addition to the light Higgs. The mass of these two fields 
is not EW, so they are {\it naturally} heavy enough to decay in
$t \bar t$. Their mass difference depends on the $\mu$ parameter and the stop masses and trilinears
in addition to $\tan \beta$\cite{deBoer:1994he}. Varying these parameters, 
for $m_A= 500$ GeV we obtain typical values of $m_H-m_A$ between $-2$ and $+10$ GeV.

\begin{figure}[H]
    \begin{center}
        {\includegraphics[scale=0.7]{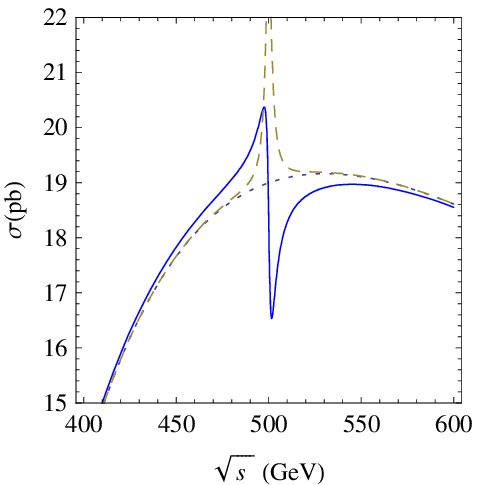}}
        {\includegraphics[scale=0.7]{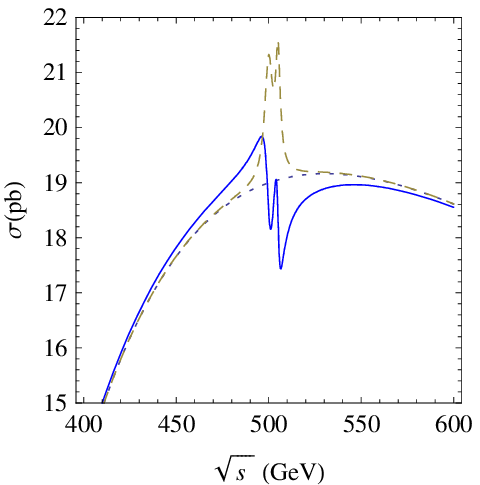}}
        \caption{\it $\sigma(gg\rightarrow t\bar t)$ for $\tan\beta=2$
and SUSY bosons of mass $m_A=m_H=500$ GeV (left)
or $m_A=500$, $m_H=505$ GeV (right). Dashes provide 
the narrow-width approximation and dots the standard model
cross section.}
\label{fig2}
    \end{center}
\end{figure}

In fig.\ref{fig2}-left we observe an average 5.5\% excess and  
8.1\% deficit in the 5 GeV intervals before and after 
$\sqrt{s}=500$ GeV, respectively. There the position of
the peaks and dips caused by
$H$ and $A$ overlap {\it constructively}. In contrast, in fig.\ref{fig2}-right
their mass difference implies a partial cancellation between 
the dip caused by $A$ and the peak of $H$. 

\begin{figure}[H]
    \begin{center}
        {\includegraphics[scale=0.7]{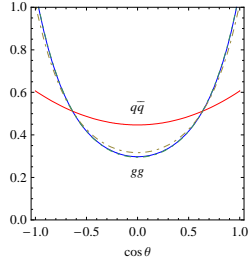}}
        \caption{\it Standard angular distribution for the $t$ quarks
from $q\bar q$ and $g g$ collisions 
at $\sqrt{s}=500$ GeV. We include (dashes) the distribution
from $g g$ at the peak and the dip of fig.\ref{fig2}-left.}
\label{fig3}
    \end{center}
\end{figure}

From fig.\ref{fig3} we argue that different
cuts could be applied to reduce the background for $t\bar t$ 
production at the LHC or even to optimize 
the contribution from $gg$ versus $q\bar q$, but not to enhance
the relative effect of the scalars on $\sigma(gg\rightarrow t\bar t)$.
\section{Little Higgs boson}
In LH models the Higgs appears as a pseudo-Goldstone
boson of a global symmetry broken spontaneously 
at the scale $f>v/\sqrt{2}=174$ GeV. The global 
symmetry introduces an extra $T$ quark
and a massive scalar singlet $r$, the {\it Higgs} 
of the symmetry broken at $f$. Once the electroweak VEV is included the 
doublet and singlet Higgses mix\cite{Barcelo:2008je,Barcelo:2007if}. 

The extra Higgs $r$ is somehow similar to the heavier scalar 
in a doublet plus singlet model, with the 
doublet component growing with $s_\theta=v/(\sqrt{2} f)$.
If $s_\theta$ is sizeable so is its coupling to the top
quark. The coupling to the extra $T$ quark is stronger, 
but if $r$ is lighter
than $2m_T$ then its main decay mode will be into $t\bar t$. Therefore, $r$
is a naturally heavy ($m_r\approx f$) but narrow scalar
resonance with large couplings to quarks and an order one
braching ratio to $t \bar t$.

In\cite{Barcelo:2010bm} we examine this case in detail. 
The results are similar to the ones
obtained for SUSY bosons of the same mass. 

\section{Signal at the LHC}

Let us now estimate the invariant mass distribution 
of $t\bar t$ events ($m_{t\bar t}$) in $pp$ collisions at the LHC. 
We will take a center of mass energy of 7 TeV and 1 fb$^{-1}$ luminosity
and we will not apply any cuts . 
At these energies the cross section $pp\rightarrow t\bar t$
is dominated by $gg$ fusion (90\%).

\begin{figure}[H]
    \begin{center}
        {\includegraphics[scale=0.6]{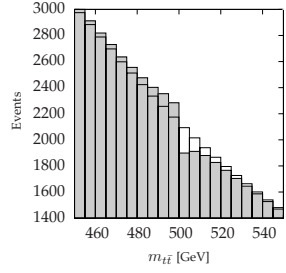}}
        \caption{\it Number of $t\bar t$ events in $pp$ collisions
at 7 TeV and 1 fb$^{-1}$ for $m_A=m_H=500$ GeV and $\tan\beta=2$ distributed in 5 GeV bins.}
\label{fig4}
    \end{center}
\end{figure}

\begin{figure}[H]
    \begin{center}
        {\includegraphics[scale=0.6]{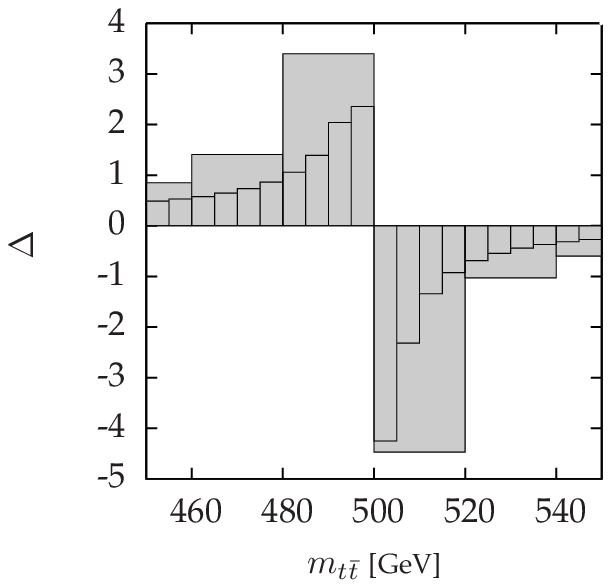}}
        {\includegraphics[scale=0.6]{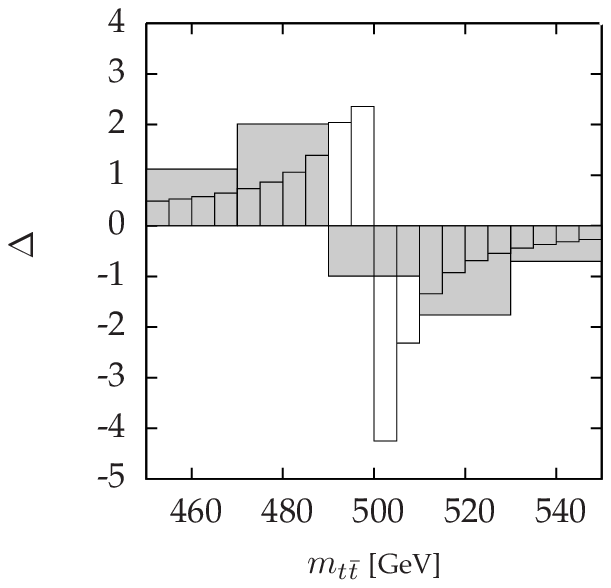}}
        \caption{\it Deviation $\Delta=(N-N_{SM})/\sqrt{N_{SM}}$ in the 
number of events  respect to the standard prediction for two
different binning ($m_A=m_H=500$ GeV and $\tan\beta=2$).}
\label{fig5}
    \end{center}
\end{figure}

In fig.\ref{fig4} we observe a 5\% excess followed by a 9\% deficit, with
smaller deviations as $m_{t\bar t}$ separates from the mass
of the extra Higgs bosons. In fig.\ref{fig5} we find that changing the 
binning is important in order to
optimize the effect. 
\section{Summary and discussion}

In models with an extended Higgs sector the extra bosons
tend to have large couplings with the top quark that imply
a sizeable one-loop production rate at hadron colliders.
If the mass of these bosons is not EW but 
comes from a new scale 
({\it e.g.}, the SUSY or the global symmetry-breaking 
scales), then they may decay predominantly  
into $t\bar t$. We have studied their effect on the $t\bar t$
invariant mass distribution at 7 TeV and  
1 fb$^{-1}$. We have considered
the deviations due to the neutral bosons $A$ and $H$ of the MSSM,
and to the scalar $r$ associated to the scale $f$ in LH
models. In all cases the interference dominates, 
invalidating the narrow-width approximation. 
\end{document}